\documentclass[lettersize,journal]{IEEEtran}
\usepackage{amsmath,amsfonts,amssymb}
\usepackage{algorithmic}
\usepackage{array}
\usepackage[caption=false,font=normalsize,labelfont=sf,textfont=sf]{subfig}
\usepackage{textcomp}
\usepackage{stfloats}
\usepackage{url}
\usepackage{cite}
\usepackage{verbatim}
\usepackage{graphicx}
\usepackage{algorithm}
\usepackage{xcolor}
\usepackage{booktabs}
\usepackage{multirow}
\usepackage{enumerate}
\usepackage{ntheorem}
\usepackage{mathrsfs}
\usepackage{psfrag}
\usepackage{tabu}
\usepackage{threeparttable}
\usepackage{graphicx}
\usepackage{adjustbox}
\usepackage{graphicx}
\usepackage{threeparttable}
\usepackage{arydshln}
\usepackage{xcolor}

\def\BibTeX{{\rm B\kern-.05em{\sc i\kern-.025em b}\kern-.08em
		T\kern-.1667em\lower.7ex\hbox{E}\kern-.125emX}}
\usepackage{balance}

\makeatletter
\renewcommand{\citepunct}{,\penalty\@m\hskip.13emplus.1emminus.1em}
\renewcommand{\citedash}{\hbox{--}\penalty\@m}

\begin{document}
	\title{Quantization Design for Deep Learning-Based \\ CSI Feedback	\vspace{-0.2cm}}
	\author{
		\IEEEauthorblockN{Manru Yin, Shengqian Han, Chenyang Yang} \\
		\IEEEauthorblockA{School of Electronics and Information Engineering, Beihang University, Beijing 100191, China\\
			Email: \{mryin, sqhan, cyyang\}@buaa.edu.cn}
		\vspace{-1.0cm}}
	
	\maketitle
	
	\begin{abstract}
		Deep learning-based autoencoders have been employed to compress and reconstruct channel state information (CSI) in frequency-division duplex systems. Practical implementations require judicious quantization of encoder outputs for digital transmission. In this paper, we propose a novel quantization module with bit allocation among encoder outputs and develop a method for joint training the module and the autoencoder. To enhance learning performance, we design a loss function that adaptively weights the quantization loss and the logarithm of reconstruction loss. Simulation results show the performance gain of the proposed method over existing~baselines.
	\end{abstract}
	\vspace{-12pt}
	
	\vspace{-10pt}
	\section{Introduction}
	
	
	Deep learning, particularly autoencoder, has shown promising for  channel state information (CSI) feedback by compressing CSI with an encoder and reconstructing it with a decoder~\cite{wen2018deep,guo2022overview,lu2020multi,ji2021channelattention,cui2022transnet}. However, early studies often overlooked the crucial aspect of quantization in practical digital communication systems~\cite{guo2022overview,lu2020multi,ji2021channelattention,cui2022transnet,wen2018deep}. Directly transmitting the raw encoder outputs, (usually a 32-bit floating-point value per output), incurs excessive overhead~\cite{guo2022overview}. Effective quantization of the encoder outputs is essential to reduce transmission~overhead.
	
	Scalar quantization, applied to each encoder output, is a widely used method in the literature~\cite{ Song2021SALDR,Chen2024CSI,guo2020convolutional,hu2024learnable,Liu2020Efficient}.
	Prior works have explored various scalar quantization techniques, including uniform quantization~\cite{Song2021SALDR}, nonlinear $\mu$-law quantization~\cite{Chen2024CSI,guo2020convolutional}, and learned nonlinear quantization~\cite{hu2024learnable,Liu2020Efficient}. Besides scalar quantization, vector quantization considers correlations among encoder outputs. In~\cite{rizzello2023user}, encoder outputs are grouped into equal-size vectors, each quantized with a vector codebook.
	Most scalar or vector quantization methods allocate bits equally across encoder outputs~\cite{ Song2021SALDR,Chen2024CSI,guo2020convolutional,hu2024learnable,Liu2020Efficient,rizzello2023user}. Recently, a novel bit allocation scheme has been proposed to enhance compression performance, where varying numbers of bits are allocated to encoder outputs based on their importance or statistical characteristics~\cite{Nerini2023Machine}.


	Since quantization is non-differential, directly incorporating it into end-to-end autoencoder training entails vanishing gradients. One way to circumvent this is to decouple the training phases: first train the autoencoder without quantization, then use the trained encoding policy to optimize the quantization codebook~\cite{yin2022deep}.
	Another way is to approximate quantization with a smoothing function during backward propagation~\cite{liang2022changeable}. This requires careful design of the smoothing function or precise adjustment of parameters, otherwise it will lead to a forward-backward propagation mismatch, degrading learning performance. {In~\cite{Mashhadi2021Distributed}, the quantization is approximated by adding noise to the encoder output during training. However, as analyzed in \cite{guo2020convolutional}, this approach may cause performance degradation during testing when ideal quantization is used.}
	An improved loss function combining weighted quantization loss and reconstruction loss was employed in~\cite{huh2023straightening}, which helps mitigate the mismatch implicitly.
	
	In this letter, we investigate a quantization scheme for deep learning-based CSI feedback. Observing that the encoder outputs have distinct dynamic ranges {through simulations}, we propose a novel bit allocation method among the encoder outputs. Unlike~\cite{Nerini2023Machine}, which optimizes bit allocation with fixed non-deep learning encoder/decoder, we develop an alternating training method to jointly learn the encoder, decoder, quantization codebook, and bit allocation. To address the aforementioned gradient mismatch issue, we design adaptive weights to combine quantization loss and reconstruction loss in the loss function, rather than fixed weights as in~\cite{huh2023straightening}. In addition, we incorporate the logarithm of the reconstruction loss, enhancing convergence and learning performance, which are often hindered by the small value of reconstruction loss. Simulation results demonstrate that the proposed method significantly enhances CSI feedback quality compared to existing~methods.

	\vspace{-8pt}
	\section{System Model}
	\vspace{-2pt}
	
	Consider a multi-input multi-output orthogonal frequency division multiplexing (MIMO-OFDM) system, where the base station (BS) with $N_t$ transmit antennas serves a single-antenna user over $\tilde{N}_c$ subcarriers. The downlink CSI is denoted as $\tilde{\mathbf{H}}\in \mathbb{C}^{\tilde{N}_c \times N_t}$. The process of CSI compression and reconstruction involves the following steps~\cite{wen2018deep,rizzello2023user}.
	
	First, {as widely considered in the literatures on the learning-based channel feedback~\cite{cui2022transnet,wen2018deep,liang2022changeable,guo2022overview}, the channels in the angle-delay domain typically exhibits significant sparsity, with most signal energy concentrated in a few dominant multipath components. Thus,} 	$\tilde{\mathbf{H}}$ is transformed into the angle-delay domain with a two-dimensional inverse discrete Fourier transform. It is then truncated to ${\mathbf{H}}\!\in \!\mathbb{C}^{{N}_c \!\times\! N_t}$, where ${N}_c \!<\!  \tilde{N}_c$ denotes the number of rows after truncation. {By selecting an appropriate truncation dimension (typically much smaller than the number of subcarriers), the information loss from truncation is very small or even negligible compared to the loss from dimension compression and quantization.} The truncated CSI  is further compressed in dimensionality by an encoder. Denote $\pmb{z} \!\in\! \mathbb{R}^{M  \!\times \!1}$ is the encoder output, and ${M}$ is the dimension of the output.
	
	Then, the encoder outputs are quantized into discrete values as $\pmb{z}^{q}$,
	where $\pmb{z}^{q}\in \mathbb{R}^{M  \times 1}$ is the quantized vector, and the quantization loss is defined as~$ \mathbb{E}_{\mathbf{H}}\big\{||\pmb{z}^q - \pmb{z}||_2^2\big\}$.
	Considering a scalar quantization policy, the $m$-th element of the encoder outputs $\pmb{z}$, denoted by $z_m$, is quantized with $B_m$ bits as
	\begin{equation}\label{fix_quan}
		z^q_{m} \!= \! w_{m,j},\ j \!=\! \text{argmin}_i\!\big\{\left(w_{m,i}  \!- \! z_m\right)^2\!, i  \!= \! 1,\ldots , 2^{B_m}\big\},
	\end{equation}
	where $z^q_{m}$ is the $m$-th element of the quantized vector $\pmb{z}^q$, $B_m$ is the number of bits allocated to the $m$-th output, and $\{w_{m,1}, \dots, w_{m,2^{B_m}}\}\triangleq \mathcal{W}_m$ form the quantization codebook for the $m$-th output. This is a non-uniform quantization when the codewords in $\mathcal{W}_m$ are not uniformly spaced.
	
	Next, the quantized vector $\pmb{z}^q$ is converted into a binary bitstream $\pmb{x}\in \{0,1\}^{MB  \times 1}$, where $B = \frac{1}{M}\sum_{m=1}^M B_m$ represents the average number of quantization bits per encoder output. The bitstream is transmitted to the BS.
	
	Finally, the BS recovers the bitstream into a discrete vector $\pmb{z}^{q}$, and then reconstructs the CSI using the decoder. Denote $\mathbf{\hat{H}}\in \mathbb{R}^{{N}_c \times N_t}$ is the reconstructed CSI. After reconstruction, {$\mathbf{\hat{H}}$ is zero-padded to make its dimension the same as the original channel, and then a two-dimensional discrete Fourier transform is applied to recover the channel in the spatial-frequency domain.}
	The reconstruction loss is defined as $\mathbb{E}_{\mathbf{H}}\big\{||\mathbf{\hat{H}} - \mathbf{{H}}||_2^2\big\}$.
	
	\section{Proposed Learning Method for Quantization}
	{As demonstrated in later simulations,}  the dynamic ranges of different elements in the encoder’s output vector differ, making equal bit allocation among these outputs suboptimal. Ideally, more bits should be allocated to the outputs with larger dynamic ranges. However, integrating bit allocation into the end-to-end training of encoder, codebook, and decoder modules is difficult due to the non-differential nature of bit allocation. To address this issue, we resort to an alternating learning approach for these four modules, consisting of two key stages: 1) given the encoder and decoder modules, we optimize bit allocation and corresponding codebook (whose dimension and values vary with bit allocation); 2) given the bit allocation, we train the encoder, codewords, and decoder modules in an end-to-end manner. These alternating steps are repeated throughout the training process, allowing for iterative refinement of these four~modules.
	
	\vspace{-10pt}
	\subsection{Design of Bit Allocation}\label{S:bit}
	
	The optimization of bit allocation aims to minimize the quantization loss $\mathbb{E}_{\mathbf{H}}\big\{||\pmb{z}^q - \pmb{z}||_2^2\big\}$ given a total bit budget of $MB$. {Learning the integer bit allocation $B_m$ is challenging, as the integer constraint is non-differentiable,} making it incompatible with gradient-based optimization. To address this, we propose an iterative bit allocation algorithm.

	
	\begin{figure*}[!h]
		\centering
		\vspace{-22pt}
		\includegraphics[width=0.90\linewidth]{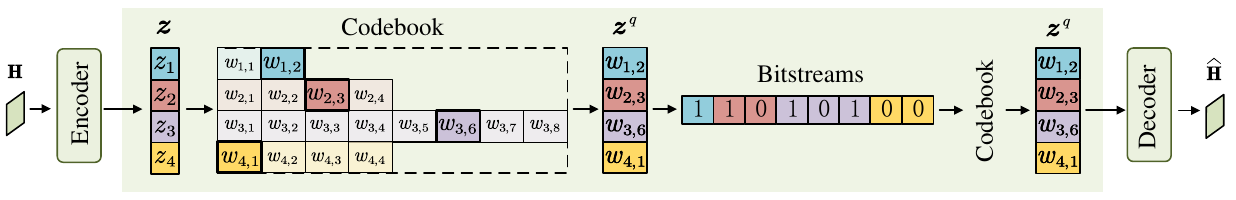}
		\vspace{-10pt}
		\caption{Illustration of the bit allocation.}
		\label{fig:Codebook}
		\vspace{-15pt}
	\end{figure*}

	In each iteration, we select two output elements: one to decrease its bit count by one and the other to increase its bit count by one, ensuring that {the total bit budget remains unchanged.}
	This bit adjustment process affects the overall quantization loss, as decreasing bits increase loss while increasing bits decrease loss.
	To minimize total quantization loss, we select elements as follows. First, we identify the element whose quantization loss increases the least when its bit count is reduced by one. Second, we identify the element whose quantization loss decreases the most when its bit count is increased by one.
	
	Specifically, denote the quantization loss of the $m$-th encoder output $z_m$ under a given number of quantization bit $B_m$ as $l_{[B_m]}=\mathbb{E}_{\mathbf{H}}\big\{(z_m^q-z_m)^2\big\}$. The value of $l_{[B_m]}$ depends on the codebook $\mathcal{W}_m$, from which $z_m^q$ is selected as shown in \eqref{fix_quan}. Based on the fixed encoder, we can generate sample set $\phi_m$ for each encoder output $z_m$, and then obtain the codebook $\mathcal{W}_m$ by using $K$-means clustering method.
	The quantization loss $l_{[B_m]}$ can be estimated by samples as $l_{[B_m]}\approx \frac{1}{|\phi_m|}\sum_{z_m\in\phi_m}\left(z_m^{q} - z_m\right)^2$,
	where $z_m^{q}\in\mathcal{W}_m$ is the quantized value (i.e., the nearest codeword) of $z_m$, and $|\phi_m|$ is the size of $\phi_m$.
	
	Similarly, we can obtain the values of $l_{[B_m- 1]}$ and $l_{[B_m+ 1]}$. The indices of the encoder outputs leading to the minimal increase and maximal decrease in quantization loss, denoted by $d$ and $a$, can be determined as
	\begin{equation}\label{dec_element}
		d = \text{argmin}_m\{l_{[B_m\!-\!1]} - l_{[B_m]}, m =1,\ldots , M\},
	\end{equation}
	\begin{equation}\label{inc_element}
		a = \text{argmax}_m\{l_{[B_m]} - l_{[B_m + 1]}, m =1,\ldots , M\}.
	\end{equation}
	
	The proposed iterative algorithm is as follows. In each iteration, the two output elements are found using \eqref{dec_element} and \eqref{inc_element}, and the bit allocation is updated by letting $B_d=B_d-1$ and $B_a =B_a+1$. This iterations are repeated until the total quantization loss of all encoder outputs remains unchanged. Since the quantization loss does not increase during the iterations, the convergence is guaranteed.
	
	{The convergence of the proposed iterative bit allocation algorithm is guaranteed. In each iteration, the bit allocation is updated such that the total quantization loss decreases or remains unchanged. Furthermore, the quantization loss is bounded below by zero. Thus, according to the Monotone Convergence Theorem~\cite{Bibby_1974}, which states that a non-increasing and bounded-below sequence of real numbers converges to its infimum, the proposed iterative algorithm converges.}
	
	\vspace{-10pt}
	\subsection{Design of Loss Function}
	With the updated bit allocation $B_m$, we optimize the encoder, decoder, and codebook in an end-to-end manner. The involved quantization process is also non-differential. A common solution is using the straight-through estimator (STE), together with incorporating quantization loss into the loss function~\cite{huh2023straightening}, where the loss function of training autoencoder is defined as
	\begin{equation}\label{loss_init}
		\frac{1}{|\varPhi|}\sum\nolimits_{\mathbf{{H}}\in\varPhi}\big( ||\mathbf{\hat{H}} - \mathbf{{H}}||_2^2+ \beta ||\pmb{z}^q - \pmb{z}||_2^2\big),
	\end{equation}
	where $\beta$ is a fixed weight, $\varPhi$ is a set consisting of samples of $\mathbf{{H}}$, and $|\varPhi|$ is the size of $\varPhi$.  The loss function for codebook is  $\frac{1}{|\varPhi|}\sum\nolimits_{\mathbf{{H}}\in\varPhi}\sum_{m=1}^M \big({z}_{m}^q - {z}_{m}\big)^2$.
	
	Choosing the weight $\beta$ is complex due to its intricate impact on reconstruction loss. A large $\beta$ prioritizes minimizing quantization loss, potentially increasing reconstruction loss. In extreme cases, a very large $\beta$ could cause the encoder to output values identical to codewords, resulting in zero quantization loss but poor reconstruction accuracy. Conversely, a small $\beta$ might insufficiently penalize quantization errors. This leads to high reconstruction errors due to gradient~mismatch from STE.
	
	We develop a novel method to adaptively adjust the weight $\beta$, based on the observation that quantization loss depends on the value being quantized and the spacing of its neighboring codewords. Small spacing between neighboring codewords (small quantization interval) generally results in low quantization loss, allowing for a small $\beta$ as a minor penalty. Conversely, large spacing necessitates a larger $\beta$ to effectively restrict quantization loss. Therefore, the value of $\beta$ should be adjusted dynamically based on the inter-codeword spacing.
	
	For an encoder output $z_m$ that is quantized as $z^q_m=w_{m,j}$ according to \eqref{fix_quan}, let $w^L_{m,j}$ be the nearest codeword greater than $w_{m,j}$, and $w^R_{m,j}$ be the nearest codeword smaller than $w_{m,j}$. $z_m$ is quantized to $w_{m,j}$ if it falls into the interval
	\begin{align}
		z_m\in[w_{m,j} - \frac{w_{m,j}-w^R_{m,j}}{2},w_{m,j} + \frac{w^L_{m,j}-w_{m,j}}{2}].
	\end{align}
	The spacing of this interval is proportional to the distance between neighboring codewords, i.e.,
	\begin{equation} \label{E:LR}
		w_{m,j}\!+\! \frac{w^L_{m,j}\!-\!w_{m,j}}{2}\!-\!\Big(w_{m,j} \!-\! \frac{w_{m,j}\!-\!w^R_{m,j}}{2} \Big)\!\propto \! w^L_{m,j} \!-\! w^R_{m,j}.
	\end{equation}
	Therefore, the weight for $z_m$ can be chosen as
	\begin{equation}\label{E:LR2}
		\beta^{\prime}_m = \beta \left(w^L_{m,j} - w^R_{m,j}\right).
	\end{equation}
	If $w_{m,j}$ is the largest codeword, set $\beta^{\prime}_m= 2\beta(w_{m,j} - w^R_{m,j})$. If $w_{m,j}$ is the smallest codeword, set $\beta^{\prime}_m= 2\beta(w^L_{m,j} - w_{m,j})$.
	
	On the other hand, most existing CSI feedback methods directly use the reconstruction loss in the loss function as in~\eqref{loss_init} \cite{wen2018deep,liang2022changeable,huh2023straightening}. However, the reconstruction loss may decrease substantially during the training process, and eventually reach an  extremely small value (even less than $10^{-7}$). This results in slow convergence due to the gradients becoming very small. To address this, we employ the logarithm of the reconstruction loss, which produces large gradients for small reconstruction loss values.
	
	Finally, the designed loss function for autoencoder is
	\begin{equation}\label{revise_loss}
		\log\!\Big(\frac{1}{|\varPhi|}\sum\limits_{\mathbf{{H}}\in\varPhi}||\mathbf{\hat{H}} - \mathbf{{H}}||_2^2\Big) +\frac{1}{|\varPhi|}\sum\limits_{\mathbf{{H}}\in\varPhi}\sum_{m=1}^M\left({\beta}^{\prime}_m \left({z}_{m}^q - {z}_{m}\right)^2\right).
	\end{equation}
	This loss function effectively balances the reconstruction and quantization losses for  efficient training.
	
	\vspace{-13pt}
	\subsection{Alternating Training}
	\vspace{-2pt}
	
	Bit allocation depends on the dynamic range of encoder outputs, which is determined by the learned encoder. Conversely, the encoder, codebook, and decoder are affected by the bit allocation, which specifies the dimension of the codebook and thus affects the learned codewords, while the encoder is forced to output values close to the learned codewords to minimize quantization loss. Therefore, bit allocation, encoder, decoder, and codebook should be jointly trained. This is realized by alternating training in the following.
	
	In the first stage of each alternating update, with a fixed bit allocation, the encoder, codebook, and decoder are trained in an end-to-end manner using the loss function defined in \eqref{revise_loss}. For each training sample, the nearest codewords $w^L_{m,j}$ and $w^R_{m,j}$ are found, and the weights $\beta^{\prime}_m$ are then computed with \eqref{E:LR2} for the sample.
	
	In the latter stage, after each training epoch, bit allocation is updated using the proposed iterative algorithm in~Sec.~\ref{S:bit}. This involves updating codewords via $K$-means clustering whenever the bit allocation changes. By alternating between these two stages, we can effectively optimize all modules.
	
	During the online inference phase, the user compresses CSI using the encoder. The encoder outputs are quantized into codewords according to the learned codebook, which are then converted to a bitstream based on the bit allocation, as illustrated in Fig.~\ref{fig:Codebook}. Upon receiving the bitstream, the BS translates it with the codebook into the quantized vector, which is then fed to the decoder for channel reconstruction.
	
	{The computational complexity of the alternating training is analyzed as follows. First, the complexity, in terms of floating point operations (FLOPs), for training the autoencoder with various architectures was provided in \cite{cong2025time}, and the results are omitted here due to lack of space. Second, the complexity of the iterative bit allocation algorithm consists of three parts: P1) A dataset for bit allocation is generated by a single forward propagation through the encoder, whose complexity can be found in~\cite{cong2025time}. P2) In the initial step of the bit allocation, for each element in the encoder's output vector, we compute its quantization loss when increasing or decreasing its bit allocation by one.
		According to~\cite{gronlund2017fast}, the complexity of K-means clustering is $\mathcal{O}(N_{\text{cluster}}|\phi_m| \!+ \!|\phi_m| \log |\phi_m|)$, where $N_{\text{cluster}}$ is the number of clusters. Therefore, the total complexity of this part is $\mathcal{O}(\sum_{m=1}^M (2^{B_{m}-1} \!+ \!2^{B_{m}+1})|\phi_m| \!+\! 2M|\phi_m| \log |\phi_m|)$.
		P3) In each iteration, the $d$-th and $a$-th elements are selected with \eqref{dec_element} and \eqref{inc_element}, whose search complexity is $\mathcal{O}(M)$. The quantization losses for these two elements are updated by increasing or decreasing one bit for each element, resulting in 4  implementations of $K$-means clustering. The iterative process continues until the total quantization loss stops decreasing.}

{From the complexity analysis, we observe that the complexity of only part P1 increases linearly with the number of antennas. The number of subcarriers does not affect the complexity, which is because channel compression is performed in the truncated delay domain rather than the frequency domain, while the dimension of channel after truncation in the delay domain is insensitive to the number of subcarriers.}
	
	
	\vspace{-10pt}
	\section{Simulations}
	\vspace{-5pt}
	\subsection{Simulation Setup}
	\vspace{-2pt}
	Consider the commonly used dataset in CSI feedback literature~\cite{liu2012cost,wen2018deep}, which is generated with the COST 2100 channel model for an indoor pico-cellular scenario. The system operates at $5.3$~GHz with $1024$ subcarriers, and users are randomly located in a 20 m $\times$ 20 m area. The BS is equipped with a 32-antenna uniform linear array (ULA). After an inverse discrete Fourier transform transformation, the first 32 rows of the channel matrix are preserved, resulting in $\mathbf{H}\in\mathbb{C}^{32\times 32}$. The training, validation, and testing sets contain $100,000$, $30,000$, and $20,000$ samples, respectively.
	The autoencoder and codebook are trained for $2000$ epochs using the ADAM optimizer with an initial learning rate of $0.001$ and exponential decay. The minibatch algorithm is used with a batch size of $200$ to compute the loss function in an iteration. For $K$-means clustering, the batch size is $2000$. The proposed method initializes bit allocation with $B_m=B$, $m=1,\ldots, M$, and sets the well-tuned $\beta$ within the range of $0.25$ to $0.01$. All learning-based methods are implemented by Pytorch.
	
	The baselines include the following methods:
	\begin{itemize}
		\item \emph{PCA}~\cite{Nerini2023Machine}: This method uses PCA for encoding instead of autoencoder. Given the obtained encoder, bit allocation is optimized for each principal component.
		\item \emph{Lloyd}~\cite{yin2022deep}: This method first trains an autoencoder \emph{without} quantization, then the codebook for quantization is obtained by the Lloyd-Max scalar quantization.
		\item \emph{Round}~\cite{liang2022changeable}: This method uses a simple rounding function for quantization (uniform quantization).
		\item \emph{Vector}~\cite{rizzello2023user}: This method divides the encoder outputs into small vectors, and quantizes each using a shared codebook, with the codewords being optimized.
	\end{itemize}
	
	The main differences between the proposed method (denoted by \emph{Proposed}) and the baselines are as follows. All methods except \emph{PCA} use an autoencoder for encoding and decoding. Both \emph{PCA} and \emph{Lloyd} first optimize encoding/decoding and then optimize quantization. The remaining methods train encoding/decoding  and quantization in an end-to-end manner. All but \emph{Round} employ non-uniform quantization with designed codebooks.
	To mitigate gradient vanishing, \emph{Round} uses a smoothing function, and \emph{Vector} uses STE and the loss function given in \eqref{loss_init}. {We will also consider several non-open-source baselines, which will be specified later.}

	
	For a fair comparison, the same autoencoder architecture is used across all compared methods. Specifically, the convolutional neural network (CNN)-based autoencoders CSINet from \cite{wen2018deep} and CSINet-Pro from \cite{liang2022changeable}, {and Transformer-based autoencoder from \cite{cui2022transnet}} are employed in the simulations. The performance metric is the normalized mean squared error (NMSE) between the reconstructed channel $\mathbf{\hat{H}}$ and the original $\mathbf{H}$, defined as
	$10\log _{10}\big(\frac{1}{|\varPhi|}\sum_{\mathbf{H}\in\varPhi}\frac{||\mathbf{\hat{H}}-\mathbf{H}||_{2}^{2}}{||\mathbf{H}||_{2}^{2}}\big)$ in~dB.
	
	\vspace{-10pt}
	\subsection{Simulation Results}
	
	Table~\ref{Tab:joint} compares \emph{Proposed} with two non-end-to-end baselines. \emph{PCA}, as a linear dimensionality reduction method, has low complexity since it does not require training. At $M=512$, the NMSE of \emph{PCA} is comparable to \emph{Lloyd}. However, the performance of \emph{PCA} degrades significantly as $M$ decreases. \emph{Proposed} shows substantial improvements over non-end-to-end \emph{PCA} and \emph{Lloyd}, even without bit allocation and adaptive weights $\beta^{\prime}_m$ (i.e., \emph{Proposed-var1}). By comparing \emph{Lloyd-log} and \emph{Proposed-var1}, both using logarithmic loss without bit allocation and adaptive weights, we can observe the gain of end-to-end training of the autoencoder and codebook. The gain increases as $B$ decreases because the impact of quantization grows as $B$ shrinks.
	
	\renewcommand{\arraystretch}{1.2}
	\begin{table}[!h]
		\centering
		\caption{Impact of end-to-end training, CSINet.}
		\vspace{-5pt}
		\label{Tab:joint}
		\begin{threeparttable}[b]
			\begin{tabular}{c|c|cccc|c|c}
				\hline
				\multicolumn{2}{c|}{$M$}          & 	\multicolumn{4}{c|}{512}      & 256       & 128              \\ \hline
				\multicolumn{2}{c|}{$B$}     & \multicolumn{1}{c}{2} & \multicolumn{1}{c}{3} & \multicolumn{1}{c}{4} & \multicolumn{1}{c|}{5}& \multicolumn{1}{c|}{4}& \multicolumn{1}{c}{4}\\ \hline
				\multicolumn{2}{c|}{PCA}& \multicolumn{1}{c}{-6.75}  & \multicolumn{1}{c}{-11.81}  & \multicolumn{1}{c}{-15.75}  &\multicolumn{1}{c|}{-17.43}  &   \multicolumn{1}{c|}{-5.05}&\multicolumn{1}{c}{0.94} \\ \hline
				\multicolumn{2}{c|}{Lloyd}& \multicolumn{1}{c}{-7.20}  & \multicolumn{1}{c}{-11.59}  & \multicolumn{1}{c}{-15.23}  &\multicolumn{1}{c|}{-17.08}  &   \multicolumn{1}{c|}{-11.86} &-8.50\\ \hline    	
				\multicolumn{2}{c|}{Lloyd-log\tnote{1}}& \multicolumn{1}{c}{-6.78}  & \multicolumn{1}{c}{-12.01}  & \multicolumn{1}{c}{-17.05}  &\multicolumn{1}{c|}{-20.55}  &   \multicolumn{1}{c|}{-13.78} &-9.45\\ \hline    		
				\multicolumn{2}{c|}{Proposed}& \multicolumn{1}{c}{-15.42}  & \multicolumn{1}{c}{-19.25}  & \multicolumn{1}{c}{ -21.90}  &\multicolumn{1}{c|}{-23.03}  &   \multicolumn{1}{c|}{-15.12}&-10.60 \\\hline
				\multicolumn{2}{c|}{Proposed-var1\tnote{2}}& \multicolumn{1}{c}{-13.14}  & \multicolumn{1}{c}{-16.86}  & \multicolumn{1}{c}{-20.29}  &\multicolumn{1}{c|}{-20.66}  &  \multicolumn{1}{c|}{ -14.15}& -10.92\\ \hline
			\end{tabular}
			\begin{tablenotes}
				\footnotesize
				\item[1]  \emph{Lloyd-log.} trains autoencoder with logarithmic loss.
				\item[2]  \emph{Proposed-var1} is a variant of \emph{Proposed} that $B_m =B$ and $\beta^{\prime}_m=\beta$.
			\end{tablenotes}
		\end{threeparttable}
		\vspace{-25pt}
	\end{table}
	
	Table~\ref{Tab:quan_design} compares the methods using end-to-end training for the autoencoder and codebook. \emph{NQ}, a non-quantized method~\cite{wen2018deep} that is enhanced with the proposed logarithmic loss, serves as a performance upper bound for methods with quantization.
	For \emph{Vector}, the codebook size grows rapidly with $L$ and $M$, making it computationally infeasible to obtain some results (marked by dashed lines). For example, with $B=3$ and $L=4$, the codebook size is $L\times 2^{LB}\! = \!2^{14}$. Note that \emph{Proposed} is not simply a special case of \emph{Vector} (with $L=1$). This is because \emph{Vector} uses a shared codebook for quantizing different vectors~\cite{rizzello2023user}, whereas \emph{Proposed} uses a distinct codebook for each encoder output. Therefore, the performance of \emph{Vector} suffers with small $L$ and $B$ due to its limited codebook size. \emph{Proposed} significantly outperforms \emph{Round} and \emph{Vector}. To isolate the impact of logarithmic loss, we also consider \emph{Proposed-var2}, which still outperforms \emph{Round} and \emph{Vector}.
	\vspace{-10pt}
	\renewcommand{\arraystretch}{1.2}
	\begin{table}[!h]
		\centering
		\caption{Impact of quantization methods, CSINet.}
		\vspace{-5pt}
		\label{Tab:quan_design}
		\begin{threeparttable}[b]
			\begin{tabular}{c|c|cccc|c}
				\hline
				\multicolumn{2}{c|}{$M$}          & 	\multicolumn{4}{c|}{512}      & 256                   \\ \hline
				\multicolumn{2}{c|}{NQ}& \multicolumn{4}{c|}{-24.73}          &     -15.77                                       \\ \hline
				\multicolumn{2}{c|}{$B$}     & \multicolumn{1}{c}{2} & \multicolumn{1}{c}{3} & \multicolumn{1}{c}{4} & \multicolumn{1}{c|}{5}& \multicolumn{1}{c}{4}\\ \hline
				\multicolumn{2}{c|}{Round}     & \multicolumn{1}{c}{-12.89}  & \multicolumn{1}{c}{-15.79}  & \multicolumn{1}{c}{-17.11}  &   \multicolumn{1}{c|}{-19.38}  &   \multicolumn{1}{c}{-12.71} \\ \hline
				\multirow{3}{*}{Vector}  & $\!\!L\!=\! 4\!$    & \multicolumn{1}{c}{-13.03}  & \multicolumn{1}{c}{-}  & \multicolumn{1}{c}{-}  & \multicolumn{1}{c|}{-}  &   \multicolumn{1}{c}{-} \\ \cline{2-7}
				&{$\!\!L\!=\! 2\!$}      & \multicolumn{1}{c}{7.9$\times10^{-5}$}  & \multicolumn{1}{c}{-14.36}  & \multicolumn{1}{c}{-16.64}  &\multicolumn{1}{c|}{-}   &   \multicolumn{1}{c}{-13.27} \\ \cline{2-7}
				&{$\!\!L\!=\! 1\!$}   & \multicolumn{1}{c}{7.9$\times10^{-5}$}  & \multicolumn{1}{c}{8.7$\times10^{-5}$}  & \multicolumn{1}{c}{-14.38}  &  \multicolumn{1}{c|}{-16.13} &   \multicolumn{1}{c}{-12.78} \\ \hline
				\multicolumn{2}{c|}{Proposed}& \multicolumn{1}{c}{-15.42}  & \multicolumn{1}{c}{-19.25}  & \multicolumn{1}{c}{ -21.90}  &\multicolumn{1}{c|}{-23.03}  &   \multicolumn{1}{c}{-15.12} \\ \hline
				\multicolumn{2}{c|}{Proposed-var2\tnote{1}}& \multicolumn{1}{c}{-14.62}  & \multicolumn{1}{c}{-17.78}  & \multicolumn{1}{c}{-19.70}  &\multicolumn{1}{c|}{-20.21}  &   \multicolumn{1}{c}{-14.26} \\ \hline
			\end{tabular}
			\begin{tablenotes}
				\footnotesize
				\item[1]  \emph{Proposed-var2} is a variant of \emph{Proposed} that is without logarithmic loss.
			\end{tablenotes}
		\end{threeparttable}
		\vspace{-18pt}
	\end{table}
	
		\begin{figure}[!h]
		\centering
		\vspace{-10pt}
		\includegraphics[width=0.8\linewidth]{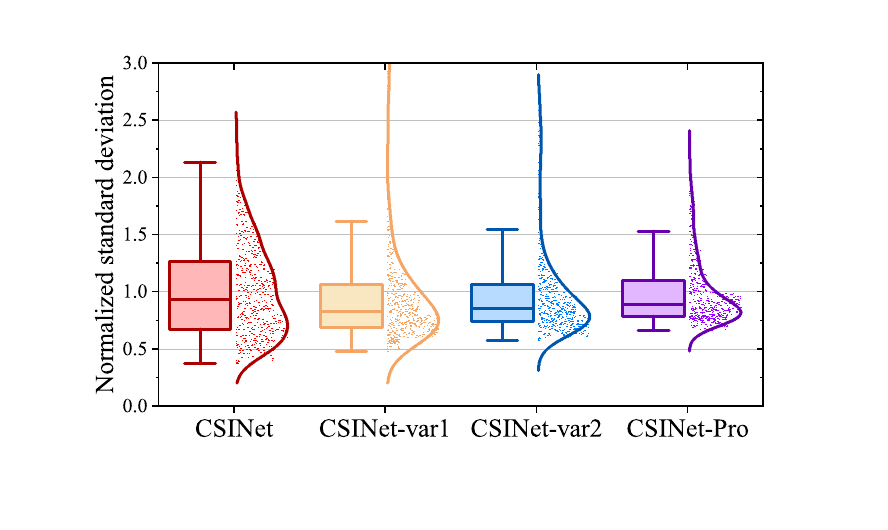}
		\vspace{-10pt}
		\caption{Normalized standard deviation for different encoders.}
		\label{fig:box}
	\end{figure}
	\vspace{-7pt}
	As previously discussed, the effectiveness of bit allocation stems from the varying dynamic ranges among encoder outputs. {Given the  difficulty of theoretically analyzing encoder behavior, we resort to simulations to quantify the dynamic ranges. For the $m$-th encoder output,} we compute the normalized standard deviation as $\frac{\sigma_m}{\bar{\sigma}}$, where $\sigma_m = \sqrt{\frac{1}{|\phi_m|}\sum_{z_m\in \phi_m}\big(z_m \!-\!\frac{1}{|\phi_m|} \sum_{z_m\in \phi_m} z_m\big)^2}$ and $\bar{\sigma}= \frac{1}{M}\sum_{m=1}^M\sigma_m$. In Fig.~\ref{fig:box},  the boxplots and fitted curves of the normalized standard deviation across the $M$ encoder outputs of \emph{NQ} are shown. The smoothed curve is fitted to the points $\frac{\sigma_m}{\bar{\sigma}}$ for $m=1,\ldots,M$.
	Four autoencoders are evaluated in Fig.~\ref{fig:box}, including CSINet, two CSINet variants (i.e., \emph{CSINet-var1} and \emph{CSINet-var2}), and CSINet-Pro, with increasing encoder scales in terms of the number of hidden convolutional layers. {{The results validate that the dynamic ranges of different output elements vary and are influenced by the encoder's architecture. Notably, encoders with fewer trainable parameters, such as CSINet, tend to produce outputs with more distinct dynamic ranges than those with more parameters,  like CSINet-Pro.}}
	
	\vspace{-10pt}
	\renewcommand{\arraystretch}{1.2}
	\begin{table}[!h]
		\centering
		\caption{Impact of dynamic environment (CSINet-Pro, $M=256$, $B=3$).}
		\vspace{-5pt}
		\label{sub_train}
		\begin{threeparttable}[b]
			\begin{tabular}{c|c|c|c|c|c|c}
				\hline
				& \multicolumn{2}{c|}{\emph{Upper}}  & \multicolumn{2}{c|}{\emph{Direct}}  & \multicolumn{2}{c}{\emph{Joint}}       \\ \hline
				\,\,\,Training set\,\, & \multirow{2}{*}{512} &\multirow{2}{*}{2048} &  \multicolumn{2}{c|}{1024}  &  \multicolumn{2}{c}{512,1024,2048} \\ \cline{4-7} \cline{1-1}
				Testing set &&   & 512& 2048& 512& 2048 \\ \hline
				Proposed	&   -19.70 & -15.08&-14.46&-8.04 &-17.60 & -13.50\\ \hline\hline\hline
				Training set & \multirow{2}{*}{In.\tnote{1}} &\multirow{2}{*}{Out.\tnote{2}} &In. &  Out. &  \multicolumn{2}{c}{In. and Out.} \\ \cline{4-7} \cline{1-1}
				Testing set & &   & Out.& In.& In.& Out. \\ \hline
				Proposed	&  -16.94 & -8.05&1.24 &-0.01& -14.12&-7.87 \\ \hline
			\end{tabular}
			\begin{tablenotes}
				\footnotesize
				\item[1-2] \emph{In.} and \emph{Out.} refer to Indoor pico-cellular and Outdoor rural scenarios from COST 2100 model.
			\end{tablenotes}
		\end{threeparttable}
		\vspace{-16pt}
	\end{table}
	
	{{Next, we discuss the application of the proposed method in dynamic environments. First, a change in the number of antennas affects the input dimension of the encoder. Thus, separate autoencoders need to be trained for different antenna configurations. The system can then select the appropriate autoencoder based on the actual antenna in use. Second, when the number of subcarriers $\tilde{N}_{c}$ changes, the channel can be truncated to the same dimension in the delay domain. Thus, a trained model can be used for different values of $\tilde{N}_{c}$, which, however, suffers from large performance degradation, as shown in the upper part of Table~\ref{sub_train}, where the method denoted as \emph{Upper} trains and tests with the same $\tilde{N}_{c}$ while the method denoted as \emph{Direct} trains the autoencoder with 1024 subcarriers and  tests with 512 or 2048 subcarriers. To address this issue, the joint training method (denoted as \emph{Joint}) uses samples from datasets with different $\tilde{N}_{c}$ for training, achieving a significant performance improvement. Third, when channel statistics change, similar results to those with changes in subcarriers can be observed. The results are shown in the lower part of Table~\ref{sub_train}, where \emph{Joint} remains effective in the two considered scenarios with different channel statistics (see the note of Table~\ref{sub_train}).}}
	
	\vspace{-15pt}
	\renewcommand{\arraystretch}{1.2}
	\begin{table}[!h]
		\centering
		\caption{Comparisons under different simulation settings.}
		\vspace{-5pt}
		\label{Tab:TransNet}
		\begin{threeparttable}[b]
			\begin{tabular}{c|c|c|c|c|c|c}
				\hline
				\textbf{$\,\,\,$TransNet$\,\,\,$} & \multicolumn{3}{c|}{Round}                          & \multicolumn{3}{c}{Proposed}                        \\ \hline
				$B$        & \multicolumn{1}{c|}{2} & \multicolumn{1}{c|}{3} & 4 & \multicolumn{1}{c|}{2} & \multicolumn{1}{c|}{3} & 4 \\ \hline
				NMSE     & \multicolumn{1}{c|}{-16.92}  & \multicolumn{1}{c|}{-19.66}  &  -21.63&\multicolumn{1}{c|}{-20.28}  & \multicolumn{1}{c|}{-23.16}  & -26.34  \\ \hline\hline\hline
				{\textbf{\cite{Mashhadi2021Distributed}}}  & \multicolumn{3}{c|}{CABAC}                          & \multicolumn{3}{c}{Proposed}                        \\ \hline
				Bit rate       & \multicolumn{1}{c|}{0.022} & 0.055  & \multicolumn{1}{c|}{0.077} & \multicolumn{1}{c|}{0.023} & \multicolumn{1}{c|}{0.035} & 0.059 \\ \hline
				NMSE    & -8.60&-11.83 &-12.45 & -18.22&-19.16&-21.86  \\ \hline\hline\hline
				{\textbf{\cite{Liu2020Efficient}}}  & \multicolumn{3}{c|}{CSIQuan}                        & \multicolumn{3}{c}{Proposed}                        \\ \hline
				$B$        & \multicolumn{1}{c|}{2} & \multicolumn{1}{c|}{3} & 4 & \multicolumn{1}{c|}{2} & \multicolumn{1}{c|}{3} & 4 \\ \hline
				NMSE     & -10.25&-12.60&-14.00& -12.52&-14.87&-15.70   \\ \hline
			\end{tabular}
		\end{threeparttable}
		\vspace{-8pt}
	\end{table}

	{Finally, in the upper part of Table~\ref{Tab:TransNet}, we show the performance gain of \emph{Proposed} when using TransNet as the autoencoder. In the middle and lower parts, we compare \emph{Proposed} with the baselines (with legends  \emph{CABAC}~\cite{Mashhadi2021Distributed} and \emph{CSIQuan}~\cite{Liu2020Efficient}), where \emph{Proposed} is simulated under the respective simulation conditions of~\cite{Liu2020Efficient} and~\cite{Mashhadi2021Distributed}. The results demonstrate the performance gain achieved by \emph{Proposed}.	}
	
	\vspace{-5pt}
	\section{Conclusion}
	This paper studied the quantization scheme for deep learning-based CSI feedback. Recognizing the varying dynamic ranges of encoder outputs, we proposed a bit allocation method across the encoder outputs and developed an alternating training method to jointly learn the encoder, decoder, codebook, and bit allocation. Furthermore, we developed a new loss function incorporating a logarithmic reconstruction loss and an adaptively weighted quantization loss, effectively enhancing the learning performance. Simulation results demonstrated that the proposed method significantly outperforms existing baselines, achieving substantial reductions in CSI reconstruction NMSE across various scenarios and autoencoder architectures.
	\vspace{-14pt}
	
	\bibliographystyle{IEEEtran}
	\bibliography{IEEEabrv,refer}
	
\end{document}